# Laser Manipulation of H⁻ Beams: a Snowmass 2022 White Paper[*]

Abdurahim Rakhman

Spallation Neutron Source, Oak Ridge National Laboratory, Oak Ridge, TN, USA 37831

Future high-power (10 – 20 MW) hadron accelerators offer many unique opportunities for fundamental science and nuclear applications. With the completion of Proton Power Upgrade (PPU) at 2.8 MW [1] and European Spallation Source (ESS) at 5 MW [2], the attention will be focused on the next high-power machine in hadron accelerator community. Charge exchange injection into a synchrotron using diamond stripper foils offers great advantage over direct proton acceleration. However, there are clear limitations associated with foil sublimation above critical beam power density and beam loss due to scattering (~10 rem/hour at 10 MW for SNS).

Over the last two decades, significant advancement of laser technology accompanied with increasing demands of H⁻ beam manipulation with lasers have enabled a wider use of lasers at such facilities. Nowadays, laser manipulations of the H⁻ beams are an important subject for the state-of-the-art high-intensity proton accelerators with many purposes. The basic principles of H⁻ beam manipulation with lasers is based on photodetachment processes by either removing a loosely bound outer electron or exciting the ground state second electron to a higher energy state to completely remove it from the H⁻ ions. The former is producing neutral H⁻ (H⁰) and therefore called H⁻ neutralization and the latter is called laser charge exchange injection or laser stripping. For the neutralization process, the energy required for photodetaching an electron is about 0.75 eV, and the interaction cross section is about $4 \times 10^{-17}$ cm² for photons of about 1.17 eV (1064 nm). Therefore, a significant fraction of the ion beam can be neutralized by focusing a 1064 nm laser beam with pulse energy on the order of 100 mJ. However, depending on the ion beam energy, detaching the second inner bound electron would require much higher laser energy and power and difficult to achieve.

Laser H⁻ beam interaction techniques such as Laser Assisted Charge Exchange (LACE) are progressing towards high power accelerator needs [3, 4, 5]. Other techniques using Laser Photo Neutralization (LPN) of H⁻ beams at low (keV), medium (MeV) and high (GeV) energies are being used for a wide range of applications such as laser-based beam control (laser notching [6], phase space sculpting [7]), non-invasive beam diagnostics [8, 9]) and beam extraction (transmutation [10], μSR source [11, 12]) in H⁻ accelerators. Future high impact accelerator applications require high repetition rate (100's MHz) intense ultrashort laser pulses bunched into short (μs – ms) bursts with slow repetition rates (10's Hz) to match accelerator bunch structure. Current laser capability is far from meeting these needs due to stringent requirements for arbitrary pulse pattern, pulse energy and pulse stability. To date, the largest challenges with all these applications are the high energy per pulse and high average laser power within the pulse pattern requirement of the accelerator. Current laser technology seems only providing one not both.

H⁻ beam manipulation with lasers has been rapidly progressing especially topics related to using neutralization to achieve beam chopping, beam halo reduction [6], beam extraction and arbitrary beam pattern generation. Concepts such as utilizing lasers to control the longitudinal emittance and reduce the beam halo for Fermilab momentum collimation has important contribution to topics for studying high power proton beam machines for the future. The concept of phase space sculpting [7] uses lasers to extract a narrow beam of neutralized hydrogen from the parent H⁻ ion beam before it gets stripped by a foil. In this concept, subsequent foil stripping and capture of protons into a storage ring generates cool proton bunches with significantly reduced emittance compared to the parent H⁻ beams. Beam extraction by arbitrary pulse patterning technique allows parasitic generation of secondary beams with many purposes such as SNS μSR project [11, 12] and SEEMS project [13]. LACE has great potential to replace the foil-based injection mechanism to realize high intensity high brightness multi-MW proton beams with minimized losses in future accelerators. By using power enhancement cavity for burst-mode lasers [14] and utilizing fiber laser and diode-pumped solid-state laser amplifier technology, we can reduce power required for full cycle laser stripping. Sequential excitation [5] and crab-crossing [15] ideas have indicated that we are moving one step closer to the practical implementation of laser stripping at SNS. Through the rapid development of laser technology, we only expect more opportunities and possibilities with H⁻ beam manipulation with lasers.

[*] This manuscript has been authored by UT-Battelle, LLC, under contract DE-AC05-00OR22725 with the US Department of Energy (DOE). The US government retains and the publisher, by accepting the article for publication, acknowledges that the US government retains a nonexclusive, paid-up, irrevocable, worldwide license to publish or reproduce the published form of this manuscript, or allow others to do so, for US government purposes. DOE will provide public access to these results of federally sponsored research in accordance with the DOE Public Access Plan (http://energy.gov/downloads/doe-public-access-plan).
†rahim@ornl.gov; phone: 1 315 391-4622; www.ornl.gov

## REFERENCES

[1] Galambos, J. *et al*., "Final Design Report Proton Power Upgrade Project", (Oak Ridge National Laboratory, ORNL/TM-2020/1570-R0, June 2020).
[2] Garoby, R., "The european spallation source design," Phys. Scripta 93, 014001 (2018).
[3] Cousineau, S., Rakhman, A., Kay, M., Aleksandrov, A., Danilov, V., Gorlov, T., Liu, Y., Plum, M., Shishlo, A., and Johnson, D., "First demonstration of laser-assisted charge exchange for microsecond duration H⁻ beams," Phys. Rev. Lett. 118, 074801 (2017).
[4] Cousineau, S., Rakhman, A., Kay, M., Aleksandrov, A., Danilov, V., Gorlov, T., Liu, Y., Long, C., Menshov, A., Plum, M., Shishlo, A., Webster, A., and Johnson, D., "High efficiency laser-assisted H⁻ charge exchange or microsecond duration beams," Phys. Rev. Accel. Beams 20, 120402 (2017).
[5] Gorlov, T., Aleksandrov, A., Cousineau, S., Liu, Y., Rakhman, A., and Shishlo, A., "Sequential excitation scheme for laser stripping for a H⁻ beam," Phys. Rev. Accel. Beams 22, 121601 (2019).
[6] Johnson, D. E., "Mebt laser notcher (chopper) for booster loss reduction," Proceedings of HB2018, Daejeon, Korea (2018).
[7] S. M. Gibson, S. E. A. and Nevay, L. J., "Laser sculpted cool proton beams," Proc. 10th Int. Particle Accelerator Conf. (IPAC'19), Melbourne, Australia (2019).
[8] Wong, J. C., Aleksandrov, A., Cousineau, S., Gorlov, T., Liu, Y., Rakhman, A., and Shishlo, A., "Laser assisted charge exchange as an atomic yardstick for proton beam energy measurement and phase probe calibration," Phys. Rev. Accel. Beams 24, 032801 (2021).
[9] Liu, Y., Long, C., and Aleksandrov, A., "Nonintrusive measurement of time-resolved emittances of 1-GeV operational hydrogen ion beam using a laser comb," Phys. Rev. Accel. Beams 23, 102806 (2020).
[10] Hayanori Takei, K. T. and Meigo, S.-I., "Low-power proton beam extraction by the bright continuous laser using the 3-MeV negative-hydrogen linac in japan proton accelerator research complex," J. of Nuclear Science and Technology 58(5) (2021).
[11] Liu, Y., Rakhman, A., Long, C. D., Liu, Y., and Williams, T. J., "Laser-assisted high-energy proton pulse extraction for feasibility study of co-located muon source at the SNS," Nucl. Instr. Meth. A 962, 163706 (2020).
[12] T.J. Williams, G. M., "Future muon source possibilities at the SNS (doi:10.2172/1364319)," (2017).
[13] Riemer, B., "SNS technical document, definition of capabilities needed for a single event effects test facility, dot/faa/tc-15/16," (2015).
[14] Rakhman, A., Notcutt, M., and Liu, Y., "Power enhancement of burst-mode ultraviolet pulses using a doubly resonant optical cavity," Opt. Lett. 40(23), 5562–5565 (2015).
[15] Aleksandrov, A., "A crab-crossing scheme for laser-ion beam applications," NAPAC'19, Lansing, Michigan, USA (2019).